\documentclass[12pt]{article}
\usepackage{amssymb}
\usepackage[latin1]{inputenc}
\usepackage{graphicx}
\newcommand{\text}{\rm}

\newcommand{\ug}{ \; = \; }

\newcommand{\bb}{\begin{equation}}
\newcommand{\ee}{\end{equation}}
\newcommand{\bega}{\begin{eqnarray}}
\newcommand{\ega}{\end{eqnarray}}
\newcommand{\begae}{\begin{eqnarray*}}
\newcommand{\egae}{\end{eqnarray*}}

\newcommand{\h}{\hspace*{4ex}}
\newcommand{\dis}{\displaystyle}

\newcommand{\om}{\omega}

\newcommand{\cent}{\centerline}
\newcommand{\vs}{\vspace*}

\begin{document}

\baselineskip 0.8cm

\begin{center}

{\large {\bf Diffraction Resistant Scalar Beams Generated by a
Parabolic Reflector and a Source of Spherical Waves}  }

\footnotetext{$^{\: (\dag)}$ Corresponding author:
mzamboni@decom.fee.unicamp.br}

\end{center}

\vs{5mm}

\cent{ Michel Zamboni-Rached\footnote{On leave from DECOM-FEEC,
University of Campinas}}

\vs{0.2 cm}

\centerline{{\em Department of Electrical and Computer Engineering
at the University of Toronto,}}

\cent{{\em Toronto, ON, Canada.}}

\vs{0.3 cm}

\cent{ Mariana Carolina de Assis }

\vs{0.2 cm}

\cent{{\em DECOM-FEEC, Universidade Estadual de Campinas, SP,
Brazil.}}

\vs{0.2 cm}

\centerline{\rm and}

\vs{0.3 cm}

\cent{ Leonardo A. Ambrosio }

\vs{0.3 cm}

\cent{{\em Department of Electrical and Computer Engineering, São
Carlos School of Engineering,}}

\cent{{\em University of São Paulo, SP, Brazil}}

\vs{0.3 cm}

{\bf Abstract  \ --} \ In this work, we propose the generation of
diffraction resistant beams by using a parabolic reflector and a
source of spherical waves positioned at a point slightly displaced
from its focus (away from the reflector). In our analysis,
considering the reflector dimensions much greater than the
wavelength, we describe the main characteristics of the resulting
beams, showing their properties of resistance to the diffraction
effects. Due to its simplicity, this method may be an interesting
alternative for the generation of long range diffraction resistant
waves.

\section{Introduction}
\h The phenomenon of diffraction causes gradual spatial broadening
of beams and pulses during propagation, and it can be a limiting
factor for any application where it is desired that the wave
maintains its transverse localization, such as in optical
tweezers, remote sensing, tissue characterization, free space
optics communications (FSO), optical atom guiding, among others
[1,9].

\h Ideal non-diffracting waves (also called localized waves) form
a set of free space solutions of the linear wave equation,
$(\nabla^2-{\partial_{ct}}^2)\Psi=0$, whose main characteristic is
the total resistance to the diffraction effects. But, for these
waves, the power flow through any plane perpendicular to the
propagation direction is infinite, thus making impossible their
experimental generation. However, it is possible to construct,
theoretically and experimentally, truncated versions of these
ideal beams[2], presenting finite power flow and resistance to the
diffraction effects for long (finite) distances.

\h In this paper we propose a simple way to obtain beams capable
of maintaining their transverse spot sizes for distances
considerably greater than those presented for the usual beams,
like the gaussian ones. For this purpose, we use a parabolic
reflector and a source of spherical waves positioned at a point
slightly displaced from its focus (away from the reflector) -
that, from now on, we will mention as the ``paraboloid setup".

\h It is important to note that we perform a simplistic analysis,
assuming that the field in question is a scalar obeying the usual
wave equation. We also assume that the source, placed at a point
slightly shifted from the focus, produces a scalar spherical
(isotropic) wave and that the whole phenomenon, including the
reflected field, possesses azimuthal symmetry. Such assumptions
may be inaccurate in practice, as the vector field character of
the case considered here is important and even if we consider this
scalar field as one of the transverse Cartesian field components,
it would hardly possess azimuthal symmetry. Even so, the reason
for which we adopted a scalar analysis here is that, in addition
to simplicity, our intention is just to provide a strong evidence
of the possibility of generating diffraction resistant beams
through the use of parabolic reflectors.

\h A more complete and rigorous analysis of the setup proposed
here, taking into account the vectorial nature of the fields, will
be addressed in a future paper.

\h In Section 2 we present some important concepts used through
the paper, like the Gaussian beam solution, the zero order Bessel
beam solution (one of the most important non-diffracting beams)
and its truncated version, the Fresnel diffraction integral, etc..

\h Section 3 is devoted to the theoretical description of the
beams generated by the paraboloid setup. We start with an
analytical and approximate description of some general properties
of the considered beams, which are then calculated through
numerical simulation of the Fresnel diffraction integral, with
examples in the optical and microwave frequency ranges. The
resulting beams are resistant to the diffraction effects for
considerable distances when compared with ordinary gaussian beams.

\h Sections 4 and 5 are devoted to the conclusions and
acknowledgments, respectively.

\h We believe that, due to its simplicity, the method presented
here can be a very interesting alternative for the generation of
long range diffraction resistant beams in optics and microwave
applications.

\section{Some important concepts}

\h Let us consider the free space and a field, $\Psi$, governed by
the wave equation\footnote{In this case, we also could think that
$\Psi$ represents the transverse cartesian component of the
electric field.}. In the paraxial approximation, a possible
monochromatic solution to this wave field is the the Gaussian beam
solution[1]:

\begin{equation}
\Psi(\rho,z,t)={\frac{2q^{2}\exp[\frac{-\rho^2}{4(q^2+iz/2k)}]}{2(q^2+iz/2k)}e^{ik(z-ct)}}
%\label{eq:5}
\label{eq:1}
\end{equation}
where $k=\omega/c$ ($c$ considered here as the light velocity) and
$(\rho, \phi, z)$ are the cylindrical coordinates.

 %%(azimuthal symmetry is assumed in Eq.(\ref{eq:1})).

\h The Gaussian beam in Eq.(\ref{eq:1}) is an ordinary beam,
susceptible to a transverse spatial broadening (diffraction). It
doubles its initial intensity spot radius, $\Delta\rho_{0}=2q$,
after a distance given by $z_{diff}=\sqrt{3}k\Delta\rho_{0}^2$
[1].

\h Now, when we make a continuous superposition of equal frequency
plane waves with the same amplitude/phase and whose wave vectors
lie on a conical surface with half opening angle $\theta$ (that we
will call ``cone angle''), the resulting wave is the well known
ideal \emph{non-diffracting} zero order Bessel beam [1,2,5,10]:

\begin{equation}
\Psi(\rho,z,t)= J_{0}(k\sin\theta \rho)e^{i k\cos\theta\, z}e^{-i
\om t} \,\, ,\label{eq2}
\end{equation}
which is an exact solution to the wave equation. The central spot
of the Bessel beam possesses a radius $\Delta\rho \approx
2.4/(k\sin\theta)$, and it is preserved during the propagation.
Actually, the entire beam transverse structure is preserved
indefinitely.

\h As it was already said, ideal non-diffracting beams carry
infinite power. This problem can be overcome, for instance, by
considering spatially truncated versions of these beams, i.e., we
consider ideal non-diffracting beam solutions truncated, on the
plane $z=0$, by a circular aperture of radius $R$.

\h In this case, the field emanated from the aperture is given (in
the paraxial approximation) by the Fresnel diffraction
integral[4,11] which, in the case of azimuthal symmetry, assumes
the form:

\begin{equation}
\Psi(\rho,z)=\frac{-ik}{z}e^{ik(z+\frac{\rho^2}{2z})}\int^{R}_{0}
\Psi_{apt}(\rho'')
e^{\frac{ik\rho''^2}{2z}}J_{0}(\frac{k\rho\rho''}{z})\rho''d\rho''
\label{fresnel}
\end{equation}
where the time harmonic dependence $e^{-i \om t}$ is implied. In
(\ref{fresnel}), $\rho''$ is the radial coordinate of the aperture
and $\Psi_{apt}(\rho'')$ is the beam on the aperture of radius $R$
localized on the initial plane $z=0$.

\h When the radius $R$ of the circular aperture is considerably
larger than the spot size of the ideal non-diffracting beam which
is being truncated, the beam emanated from the aperture (i.e, the
truncated beam) will be able to resist the diffraction effects for
long (finite) distances.

\h Figure 1 shows a comparison between a Gaussian and a truncated
Bessel beam, both with the same initial spot size. The caption is
self explanatory.

\h Here we define ``diffraction resistant beams'' as finite power
beams that can maintain their spot size for long distances when
compared to usual beams. The truncated Bessel beam is an example,
but there are many other beams with this characteristic[12-16].

\h In the following section we present a new way for generating
diffraction resistant beams by using a parabolic reflector and a
source of spherical waves positioned at a point slightly displaced
from its focus (away from the reflector).

\begin{figure}[htbp]
\centering \fbox{\includegraphics[width=\linewidth]{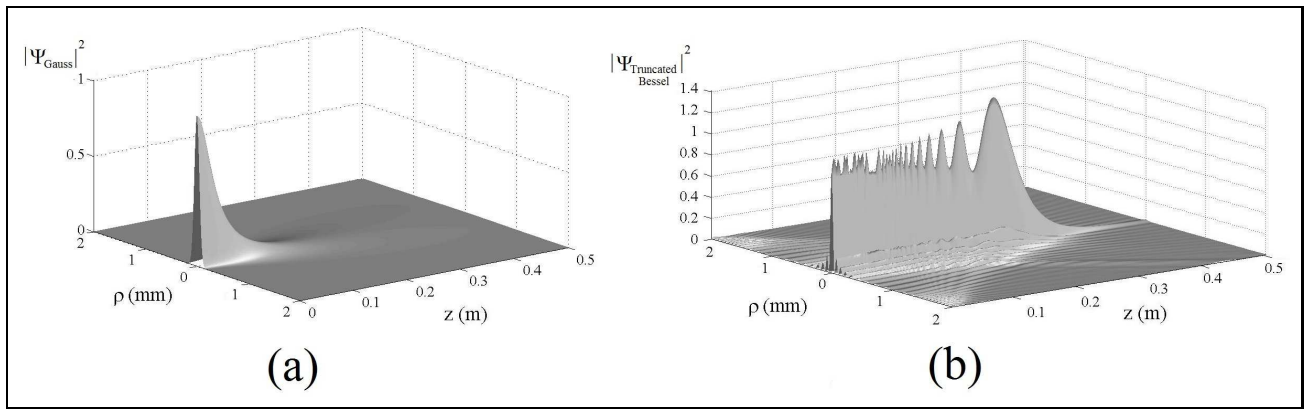}}
\caption{Comparison of (a) a Gaussian and (b) a Bessel beam
    truncated by an aperture of 2mm. Both beams possess $\lambda=850$nm and spot radius of $68\mu$m.
    The Gaussian beam intensity doubles its initial transverse width after 6 cm, and
after 10 cm its intensity decays by a factor of 10. By contrast,
the truncated Bessel beam keeps its transverse shape approximately
up to a distance of 33 cm.}
\label{fig1}
\end{figure}

\section{Generation of diffraction resistant beams by using a
parabolic reflector}

\h It is well known that in the case of a parabolic reflector,
every ray originating from its focus will be reflected in a
direction parallel to its axis. Now, if the rays originate from a
point slightly displaced away from the reflector, they will
converge (after reflection) to its axis, more specifically, they
will converge along a focal line[5,7].

\h As a focal line is a common characteristic of diffraction
resistant beams, we are tempted to ask if a parabolic reflector of
dimensions much greater than the wavelength considered and with a
source of spherical waves positioned at a point slightly displaced
away from its focus is capable to generate a diffraction resistant
beam within the extended focus range.

\h We shall analyze the paraboloid setup in three steps. In the
first one we use simple ray optics to quantify the range of the
focal line. In the second step, we proceed with a heuristic and
approximate description of the beam's transverse structure. In
spite of the theoretical development at this stage being very
approximative, it yields important information about the spot size
evolution of the resulting beams, making more understandable the
circumstances in which they would present diffraction resistance
characteristics.

\h In the third step, subsection 3.3, we use the Fresnel
diffraction integral to describe the beams emanated from the
paraboloid setup. Examples in optical and microwave frequencies
are considered and comparisons with Gaussian beams with the same
spot sizes show that these new beams really present resistance to
the diffraction effects for long distances.

\subsection{Geometric considerations}

\h Let us consider a parabolic reflector described in cylindrical
coordinates by the equation

\bb z \ug a \rho^2 \,\, , \label{parabola} \ee
with $0 \leq z \leq
z_p$, being $ z_p \gtrapprox z_f $ where $z_f = 1/4a$ is the
paraboloid focus position.

\h In the setup proposed here, the source of spherical waves is
positioned at $z=z_p$ and the following assumptions are made:

\begin{itemize}
    \item {the dimensions of the parabolic reflector are much greater than the wavelength considered}
    \item {the position $z=z_p \gtrapprox z_f$ of the source of spherical waves is such
    that

    \bb 0< \frac{z_p - z_f}{z_{f}} <<1 \,\, \label{cond1}\ee

    %%\bb 0< \frac{z_p - z_f}{z_{p \atop f}} <<1 \,\, \label{cond1}\ee
}
\end{itemize}

\begin{figure}[htbp]
\centering \fbox{\includegraphics[width=\linewidth]{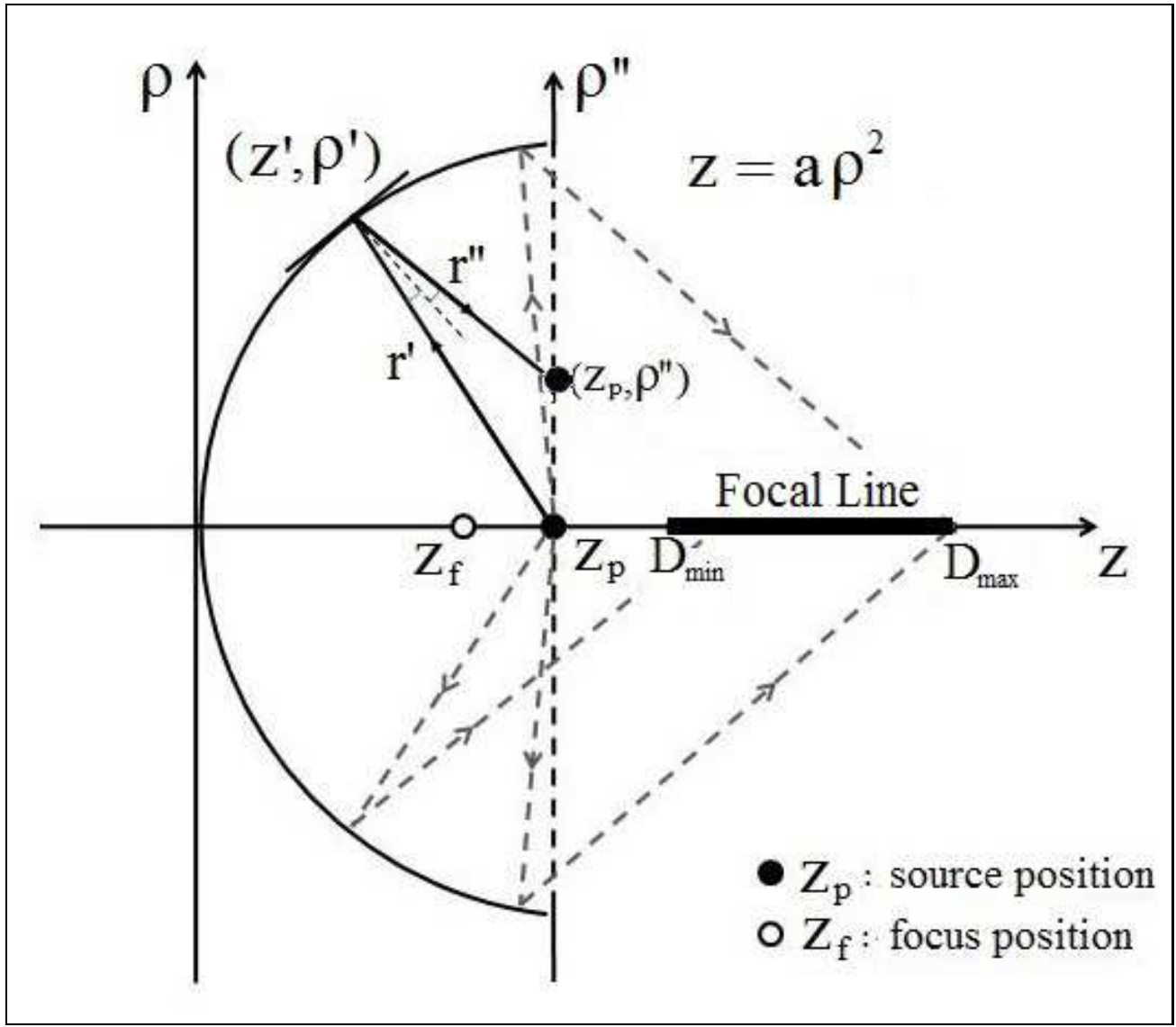}}
\caption{Paraboloid setup: a parabolic reflector with focus in
$z=z_f$ and a source of spherical waves positioned at $z=z_p$,
with $z_p \gtrapprox z_f$.} \label{fig2}
\end{figure}

Figure \ref{fig2} shows the schematics of the paraboloid setup.

\h Now, we can use geometric optics to evaluate the line-focus
length created by the rays associated with a spherical wave
positioned at $z=z_p$. The calculations are based on
Fig.\ref{fig2} and do not present difficulties[5,7]. The focal
line occurs within $D_{min} \leq z \leq D_{max} $, where

\begin{equation}
D_{min}=\frac{z_p z_f}{(z_p-z_f)} \label{dmin}
\end{equation}

\begin{equation}
D_{max}= \frac{z_p^2 + 3z_p z_f}{z_p - z_f} = z_p+4D_{min}
\label{dmax}
\end{equation} \\

\h Therefore, the line-focus length $Z_{lfl}$ can be written as:

\begin{equation}
Z_{lfl} = D_{max} - D_{min}= z_p + 3D_{min} \approx 3D_{min}
\label{zfd}
\end{equation}

\h We expect that within the range of the extended focus the
resulting beam presents resistance to the diffraction effects.
Using plausible assumptions based on geometrical optics and on a
spectral characteristic of the zero order Bessel beam, a heuristic
derivation about the transverse field pattern of the beams in
question will be presented in the next section. From that, we can
get important information about the spot size, diffraction
resistance length (field depth), etc..

\subsection{Heuristic description of the transverse field pattern of the beams generated by the
paraboloid setup}

\h As we have already said, in this work we have adopted a
simplistic analysis, assuming that the field in question,
$\Psi(\rho,\phi,z,t)$, is a scalar obeying the usual  wave
equation.

\begin{figure}[htbp]
\centering \fbox{\includegraphics[width=\linewidth]{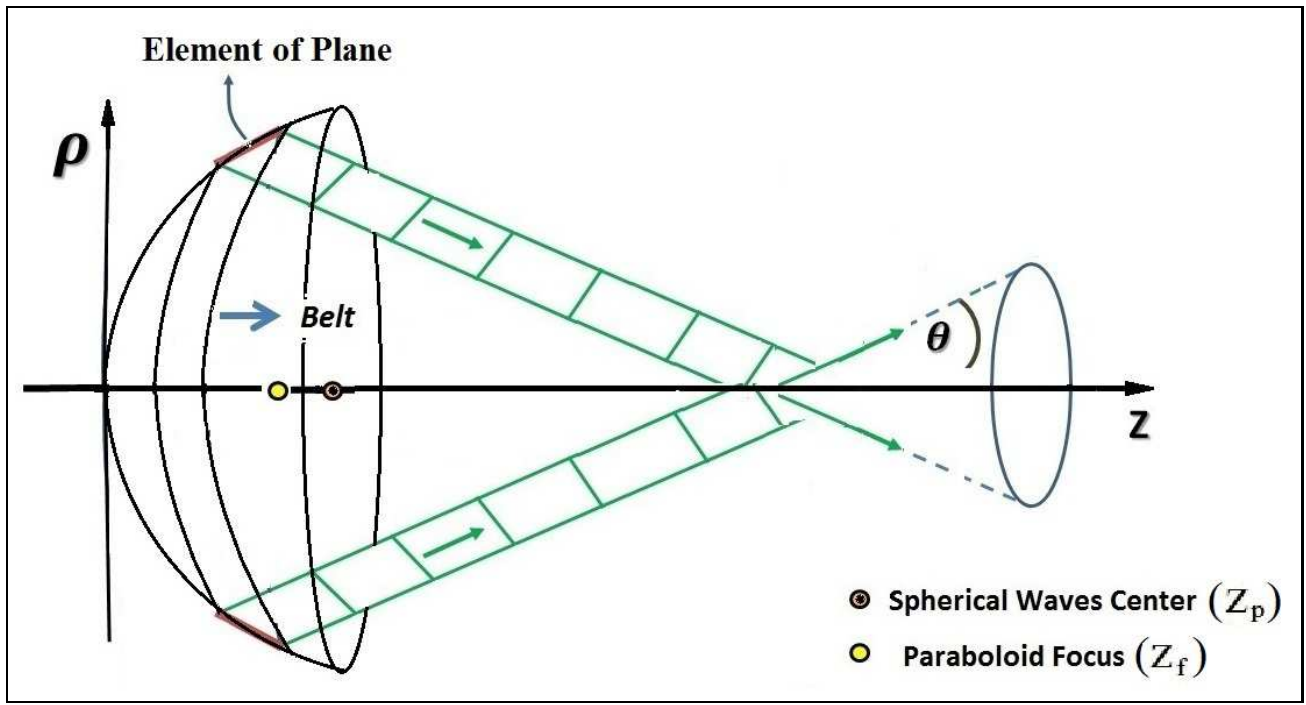}}
\caption{Discretization of the parabolic reflector in belts and
the discretization of
    these belts in elements of plane whose dimensions are taken as being much greater
    than the wavelength of the source. Each ``portion" of the spherical
    wave that reaches one of these elements of plane is reflected as a ``portion" of plane
    wave that, we assume, does not suffer considerable diffraction until it crosses the
    $z$-axis. This process results, on each $z$-axis point within the extended focus range,
    in a superposition of ``portions'' of plane waves whose wavevectors lie over a conical surface.} \label{fig3}
\end{figure}

\h Let us consider, as already said, a parabolic reflector of
dimensions much greater than the wavelength used in the source of
the setup. We can imagine a discretization of the paraboloid in
belts, which, in turn, are discretized by elements of planes with
dimensions much greater than the wavelength of the source of
spherical waves [5,7]. See Fig.\ref{fig3}.

\h Thus, we can say that the portions of the spherical wave
originated from $(\rho=0,z=z_{p})$ - the ``spherical wave center"
- that arrive at each one of these planes will be reflected as
plane waves with the dimensions of the respective incident planes.
So, each point of the $z$ axis, in the extended focus range, will
be achieved by a set of ``portions" of plane waves from the
respective reflector belt\footnote{Each belt of radius $\rho'$ on
the reflector corresponds to a $z$ point over the axis, according
to the reflected rays, as shown in Fig.\ref{fig3}.}.

\h By symmetry, the wave vectors of the set of plane wave
``portions" arriving at a point on the focal line will be on the
surface of a cone whose half opening angle is given by $\theta$
(see Fig.\ref{fig3}). In a very approximate way, we can say that
this superposition will produce a Bessel beam with cone angle
$\theta$, which will propagate by a small interval in $z$ and then
``replaced" by the Bessel beam of the next belt, and so on, until
the position $D_{max}$, which marks the end of the focal line.

\h By using geometric optics, it can be shown that

\begin{equation}
\sin\theta \ug \frac{1}{\dis{\sqrt{\left(\frac{z-\frac{\eta^{2}}{4
z_f}}{\eta}\right)^{2}+1}}} \,\, , \label{eq8}
\end{equation}
where

\begin{equation}
\eta \ug
2z_f\,\sqrt{-1+\sqrt{1-\frac{z_p}{z_f}+\frac{z_p-z_f}{z_f^2}\,z\,}\,}
\label{eq9}
\end{equation}
and $D_{min} < z \leq D_{max}$.

\h According to these general considerations and noting that the
intuitive scheme above does not provide any information about the
beam amplitude on different z positions, we conjecture that within
the focal line and for points near the z-axis, the resulting beam
can be approximately written as $\Psi(\rho,z,t) \approx A(z)
J_0(k\sin\theta \rho) \exp(i\, k \cos\theta \, z) \exp(-i \om t)$,
where $A(z)$ is an (unknown) amplitude function depending on $z$.
In this way, we can get a very useful information about the
transverse behavior of the resulting beam, for $D_{min} < z \leq
D_{max}$ and for points near the z-axis, by considering the ratio
$\Psi(\rho,z,t) / \Psi(\rho=0,z,t)$ which, in this approach, is
given by

\begin{equation}
\dis{\frac{\Psi(\rho,z,t)}{\Psi(\rho=0,z,t)}} \approx J_{0}(k \sin
\theta \rho) \label{psinorm}
\end{equation}
where it should be noted that $\theta$ is not constant but
dependent on $z$, according to Eqs.(\ref{eq8},\ref{eq9}).

\h Equation (\ref{psinorm}) can furnish important information
about the diffraction resistance characteristics of the beams
generated from this setup. Specifically, the spot size radius --
$\Delta\rho$ -- of the resulting beam can be estimated as

\bb \Delta\rho \approx \frac{2.4}{k\sin\theta} \,\, , \label{spot}
\ee
which is clearly dependent of $z$ due to
Eqs.(\ref{eq8},\ref{eq9}).

\h The expression for the spot radius is complicated, but many
simplifications can be done in the cases where the relation
(\ref{cond1}) is satisfied and, as we have already said, these are
the cases of interest here. The following results of this
subsection are obtained assuming relation (\ref{cond1}).

\h It is possible to show that the minimum value of $\Delta\rho$,
the beam waist, is approximately given by

\bb \Delta\rho_{min}  \approx 0.6 \frac{z_p}{z_p - z_f}\lambda
\label{rhomin} \ee
and occurs at $z=z_{bw}$, with

\bb z_{bw} \approx \frac{16}{9}\frac{z_f^2}{z_p - z_f} \approx
1.77 D_{min} \label{zbw} \ee

\h Within the extended focus range, the beam spot radius, after
$z=z_{bw}$, increases until reaching its maximum value at $z =
D_{max}$, with $D_{max}$ given by Eq.(\ref{dmax}). By using this
value of $z$ into Eq.(\ref{spot}) we get

\bb \Delta\rho_{max}  \approx 0.38 \frac{z_p + z_f}{z_p -
z_f}\lambda  \label{rhomax} \ee

\h So, after the distance

\bb Z = D_{max} - z_{bw} \approx 2.22 D_{min} = 2.22 \frac{z_p
z_f}{z_p - z_f} = 0.6 z_f \Delta\rho_{min} k \,\, , \label{Z} \ee
we have that

\bb \frac{\Delta\rho_{max}}{\Delta\rho_{min}} \approx 1.26  \ee

\h We consider\footnote{It is important to note that in
Eq.(\ref{Z}) for the field depth, the beam waist
$\Delta\rho_{min}$ also depends on $z_f$ -- see Eq.(\ref{rhomin})}
$Z$, Eq.(\ref{Z}), as the field depth of the resulting beam. After
this distance, its spot radius becomes $1.26$ times greater than
the beam waist, which occurs at $z = z_{bw}$.

\h This field depth can be considerably greater than the
diffraction distance\footnote{The distance wherein the Gaussian
beam intensity doubles the value of its waist.} (Rayleigh
distance) of the Gaussian beam, given by $Z_{Gauss} =
\sqrt{3}\Delta\rho_{Gauss}^2 k$, where $\Delta\rho_{Gauss}$ is the
Gaussian beam's waist. To see this, let us consider a situation
where the waists of a Gaussian beam and of a beam generated by the
paraboloid setup are equal, i.e., $\Delta\rho_{Gauss} =
\Delta\rho_{min}$. In this case we have:

\bb \frac{Z}{Z_{Gauss}} \approx 0.35 \frac{z_f}{\Delta\rho_{min}}
\,\, , \label{razao} \ee
and it is not difficult to see that the
above ratio may reach values significantly greater than unity.

\h For instance, let us assume a parabolic reflector whose focus
is localized at $z_f = 2$cm and a spherical wave source, with
$\lambda=850$nm, located at a distance of $8\lambda$ away from the
focus, i.e., $z_p = z_f + 8\lambda$. In this case, using
Eq.(\ref{dmin}) and Eq.(\ref{dmax}), we obtain the extended focus
occurring between $D_{min}=58.8$m and $D_{max}=235.4$m,
respectively. Yet, according to our approximative equations, the
resulting beam waist will be $\Delta\rho_{min} \approx 1.5$mm,
localized at $z_{bw} \approx 104$m and, from there, after a
distance of $Z = 131$m (the field depth), the spot radius will be
$\Delta\rho_{max} \approx 1.9$mm, just $1.26$ times greater. A
Gaussian beam with the same wavelength would double this same
waist after a distance $Z_{Gauss} \approx 29$m.

\h Figures 4 and 5 show, respectively, the evolutions of the
beam's transverse pattern, according Eq.(\ref{psinorm}) and of the
beam's spot radius, according Eq.(\ref{spot}), both complemented
by eqs.(\ref{eq8},\ref{eq9}).

\begin{figure}[htbp]
\centering \fbox{\includegraphics[width=\linewidth]{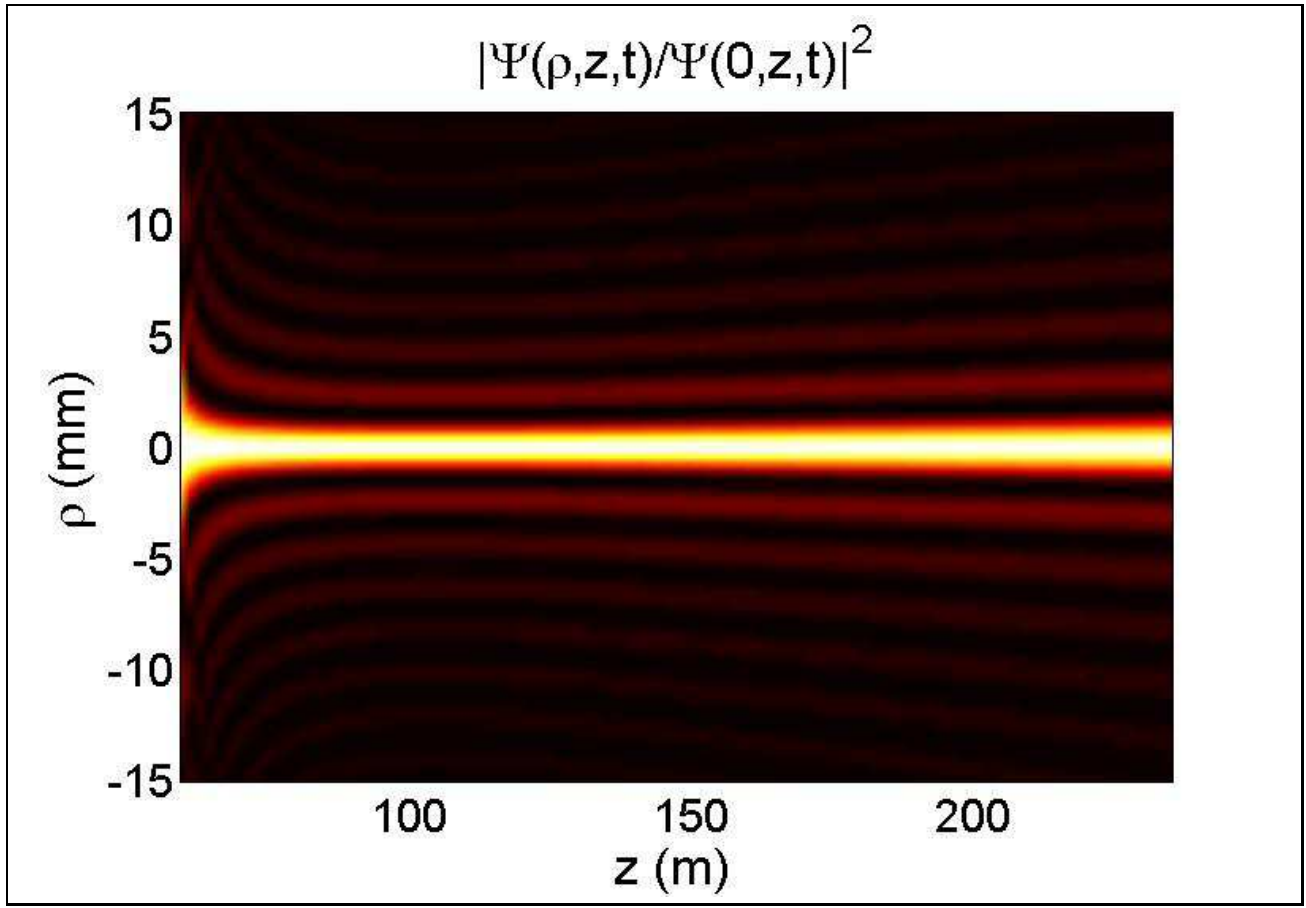}}
\caption{Evolution of the beam transverse
pattern,$\Psi(\rho,z,t)/\Psi(\rho=0,z,t)$, according
Eq.(\ref{psinorm}), with eqs.(\ref{eq8},\ref{eq9}), considering a
source of spherical waves with $\lambda=850$nm, displaced
$8\lambda$ away from a parabolic reflector whose focus is located
at $z_f=2$cm.} \label{fig4}
\end{figure}

\begin{figure}[htbp]
\centering \fbox{\includegraphics[width=\linewidth]{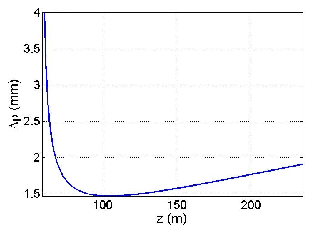}}
\caption{Variation of the spot radius within the extended focus
range, according to Eq.(\ref{spot}),
    considering a source of spherical waves with $\lambda=850$nm, displaced $8\lambda$ away from a
    parabolic reflector focus which is located at $z_f=2$cm.}
\label{fig5}
\end{figure}

\h From these figures, we can see that the beam has a diffraction
resistance behavior for a distance even greater than $Z = 131$m,
actually for a distance approximately equal to the line-focus
length $Z_{lfl} = 176$m.

\h While in this section we have presented a heuristic description
of the beam's transverse evolution based on Eq.(\ref{psinorm}), in
the following section the full resulting beam emanated from the
paraboloid setup is obtained in a more rigorous way by using the
wave theory. More specifically, we first evaluate the field on the
paraboloid aperture and from it the emanated field is calculated
through the Fresnel diffraction integral. Some examples are
presented.

\subsection{Description of the beams resulting from the paraboloid setup by using wave optics}

\h Let us consider the parabolic setup represented by
Fig.\ref{fig2}, where a spherical wave source is located at
$(\rho=0,z=z_p)$, with $z_p \gtrapprox z_f=1/4a$. The wave
produced by this source is written as:

\begin{equation}
\Psi_{sph}=\frac{e^{ikr}}{r} \label{eq12}
\end{equation}
where $r$ is the distance from the source to the point in
question.

\h So, the incident wave on the paraboloid reflector, represented
by the coordinates $(\rho',\phi',z')$ with $z'=a\rho'^2$, is given
by

\begin{equation}
\Psi_{i}=\frac{e^{ikr'}}{r'} \label{psii}
\end{equation}
where

\begin{equation}
r'=\sqrt{(z_p-z')^2+\rho'^2} \label{r'}
\end{equation}

\h We will use this incident wave to evaluate, approximately, the
field on the paraboloid aperture, that is, on the circular
aperture of radius $R = \sqrt{z_p/a}$. To achieve this, first we
note that a ray originating at the source position and reflected
at the paraboloid point $(\rho',\phi',z')$, will travel a distance
$r''$, with

\begin{equation}
r''=\sqrt{(z_p-z')^2+(\rho'-\rho'')^2} \,\, , \label{r''}
\end{equation}
and reach the plane $z=z_p$ (where the paraboloid aperture is) at
the point $(\rho'',\phi'')$, where $\phi'' = \phi'$ and

\bb \rho'' \ug \frac{4z_f\,(z_p-z_f)(4z_pz_f -
\rho'^2)\rho'}{(\rho'^2 - 4z_pz_f)(\rho'^2-4z_f^2)+16\rho'^2z_f^2}
+\rho' \,\, . \label{rho''} \ee

\h Based on the assumption that the paraboloid dimensions are much
greater than the wavelength, the next step is to consider that
each little portion of the incident spherical wave $\Psi_i$,
Eq.(\ref{psii}), will be reflected as a plane wave, in such way
that the reflected field on the paraboloid aperture, $\Psi_{apt}$,
can be approximately written as

\begin{equation}
\Psi_{apt}=-\Psi_{i}e^{ikr''}=-\frac{e^{ik(r'+r'')}}{r'} \,\, ,
\label{psiapt}
\end{equation}
where $r'$ and $r''$ are given by eqs.(\ref{r'},\ref{r''}).

\h Now, with the knowledge of the reflected field on the
paraboloid aperture, we can evaluate the emanated field by using
the Fresnel diffraction integral given by Eq.(\ref{fresnel}).

\h It is important to notice that this diffraction integral is
performed over the aperture's radial coordinate, $\rho''$, and so
the $\Psi_{apt}$, Eq.(\ref{psiapt}), should be written in terms of
this variable by inverting the relation (\ref{rho''}). Such
inversion is a difficult task, therefore we choose to change the
variable of integration in Eq.(\ref{fresnel}) from $\rho''$ to
$\rho'$, by using directly Eq.(\ref{rho''}).

\h With this, the resulting beam emanated from the paraboloid
setup can be obtained by numerically solving the Fresnel
diffraction integral over the variable $\rho'$:

\bb
 \begin{array}{clr}

 \Psi(\rho,z) \ug &
\dis{\frac{ik}{z-z_p}}e^{ik[(z-z_p)+\frac{\rho^2}{2(z-z_p)}]}  \\

\\

& \times \dis{\int^{R}_{0}}
\frac{e^{ik[r'(\rho')+r''(\rho')]}}{r'(\rho')}
e^{\frac{ik[\rho''(\rho')]^2}{2(z-z_p)}}J_{0}\left(\frac{k\,\rho\,\rho''(\rho')}{z-z_p}\right)\rho''(\rho')\frac{d\rho''}{d\rho'}
d\rho' \,\, , \end{array} \label{fresnel2} \ee
where the functions
of $\rho'$: $\rho''(\rho')$ and $d\rho''/d\rho'$ are obtained from
Eq.(\ref{rho''}), $r'(\rho')$ and $r''(\rho')$ are given by
eqs.(\ref{r'}) and (\ref{r''}), the last one complemented by
Eq.(\ref{rho''}).

\h Now that we have the integral solution for the resulting beam,
we are going to applied it to a few examples, which will confirm
the beam's diffraction resistance properties as well the
predictions of the previous sections/subsections.

\emph{\textbf{First example:}}

\h Let us use Eq.(\ref{fresnel2}) for describing the generation of
a beam with $\lambda = 850$ nm by the paraboloid setup whose focus
$z_f = 2$ cm. The source of spherical waves is located $8\lambda$
away from the focus, i.e., $z_p = z_f + 8\lambda$, which implies
an aperture of radius $R = 4$cm.

\h With these parameters, our approximative equations
(\ref{dmin},\ref{dmax},\ref{zfd},\ref{spot},\ref{rhomin},\ref{zbw},\ref{rhomax})
predict: i) a line-focus length $Z_{lfl}=175.5$m, inside the
interval $D_{min}=58.8\text{m} \leq z \leq D_{max}=235.4$m; ii) a
beam waist of $\Delta\rho_{min} \approx 1.5$mm at $z_{bw} \approx
104$m, with a diffraction resistance behavior for (at least) a
distance of $Z \approx 131$m.

\h The resulting beam is evaluated by a numerical simulation of
the diffraction integral, Eq.(\ref{fresnel2}) and the results are
shown in Figures 6 and 7.

\h Figure 6a shows the beam intensity and Fig.6b its orthogonal
projection. We can see that the extended focus occurs as
predicted, but due to the intensity variation along the distance,
the details about the spot size behavior are not very clear,
although it is clear its resistance to the diffraction effects
along the extended focus range.

\h Now, to get more information about the transverse beam
behavior, Fig.7a shows the intensity ratio
$|\Psi(\rho,z,t)/\Psi(\rho=0,z,t)|^2$ for the resulting beam
starting from $z=D_{min}=58.8\text{m}$. The evolution of the spot
size within the extended focus range $D_{min} < z \leq D_{max}$ is
satisfactorily described by Eq.(\ref{spot}) with
eqs.(\ref{eq8},\ref{eq9}), as we can see by comparing Figures 6
and 7a. We can clearly see that the size and position of the beam
waist agree with the predicted values and it is also very clear
that the beam is very resistant to the diffraction effects until
the distance $Z \approx 131$m. Actually its central spot suffers
just small variations for distances greater than the predicted
field depth $Z$, but it is also clear that after that the beam's
lateral lobes become much more pronounced\footnote{It is very
interesting that the beam central spot continues to suffer just
small variations after the predicted field depth, even if the side
lobes become more pronounced.}.

\begin{figure}[htbp]
\centering \fbox{\includegraphics[width=\linewidth]{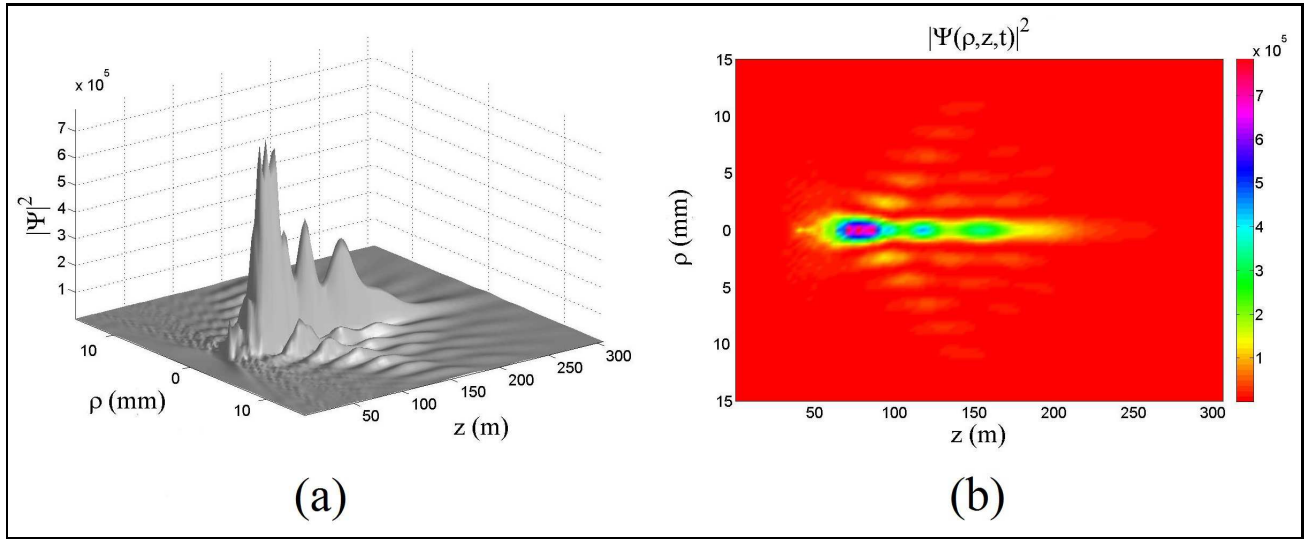}}
\caption{(a)Evolution of the beam intensity resulting from the
    paraboloid setup when $\lambda = 850$nm, $z_f = 2$cm and $z_p = z_f +
    8\lambda$; (b) The orthogonal projection of the resulting beam
    intensity. We can see, as predicted, that the beam possesses a extended
    focus range in which it presents resistance to the diffraction effects.} \label{fig6}
\end{figure}

\begin{figure}[htbp]
\centering \fbox{\includegraphics[width=\linewidth]{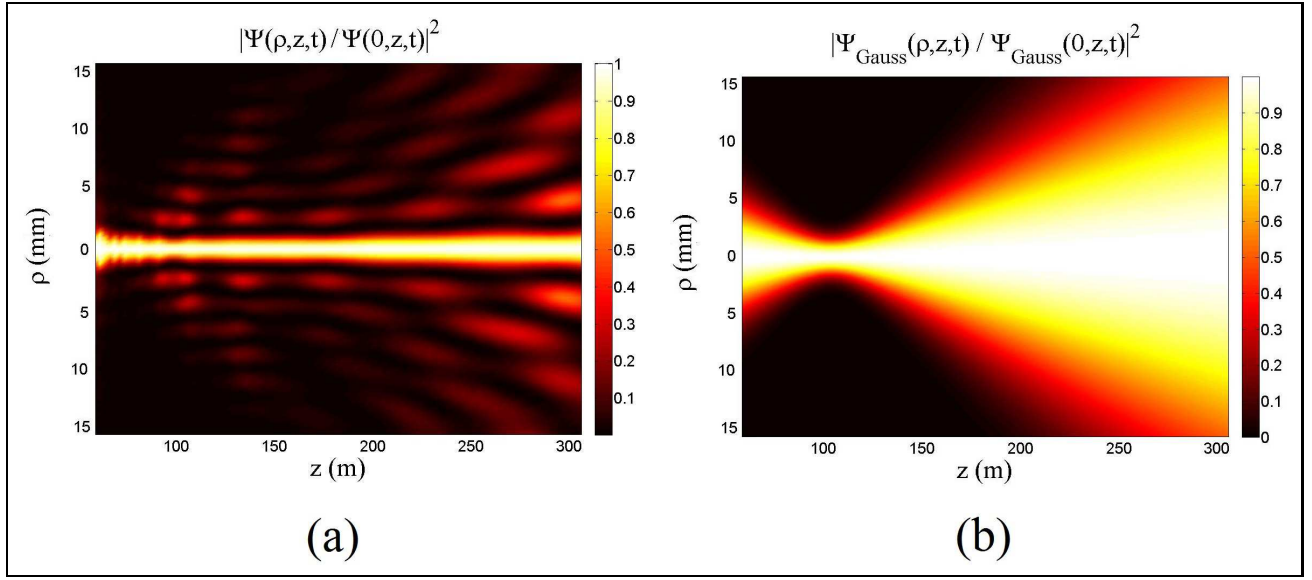}}
\caption{(a)The intensity ratio
$|\Psi(\rho,z,t)/\Psi(\rho=0,z,t)|^2$ for the resulting beam
starting from $z=D_{min}=58.8\text{m}$. We can see that the size
and position of the beam waist agree with the predicted values and
it is also very clear that the beam is very resistant to the
diffraction effects for distances even greater than $Z \approx
131$m; (b) The same kind of intensity ratio for a Gaussian beam
with the same waist at the same position. After $29$m this beam
doubles its intensity spot size.} \label{fig7}
\end{figure}

\h Figure 7b shows the same kind of intensity ratio for a Gaussian
beam with the same waist at the same position. After $29$m this
beam doubles its intensity spot size.

\emph{\textbf{Second example:}}

\h Now, let us consider our paraboloid setup to generate a beam of
wavelength $\lambda = 1$cm. In this case we choose the focus $z_f
= 0.5$m. The source of spherical waves is located at $z_p =
0.525$m, which implies an aperture of radius $R \approx 1$m.

\h For this configuration our approximative equations
(\ref{dmin},\ref{dmax},\ref{zfd},\ref{spot},\ref{rhomin},\ref{zbw},\ref{rhomax})
predict: i) a line-focus length $Z_{lfl}=32$m, inside the interval
$D_{min}=10.5\text{m} \leq z \leq D_{max}=42.5$m; ii) a beam waist
of $\Delta\rho_{min} \approx 12.6$cm at $z_{bw} \approx 18.6$m,
with a diffraction resistance behavior for (at least) a distance
of $Z \approx 24$m.

\h We evaluated the resulting beam by the numerical simulation of
the diffraction integral, Eq.(\ref{fresnel2}), and the results are
shown in Figures 8 and 9.

\begin{figure}[htbp]
\centering \fbox{\includegraphics[width=\linewidth]{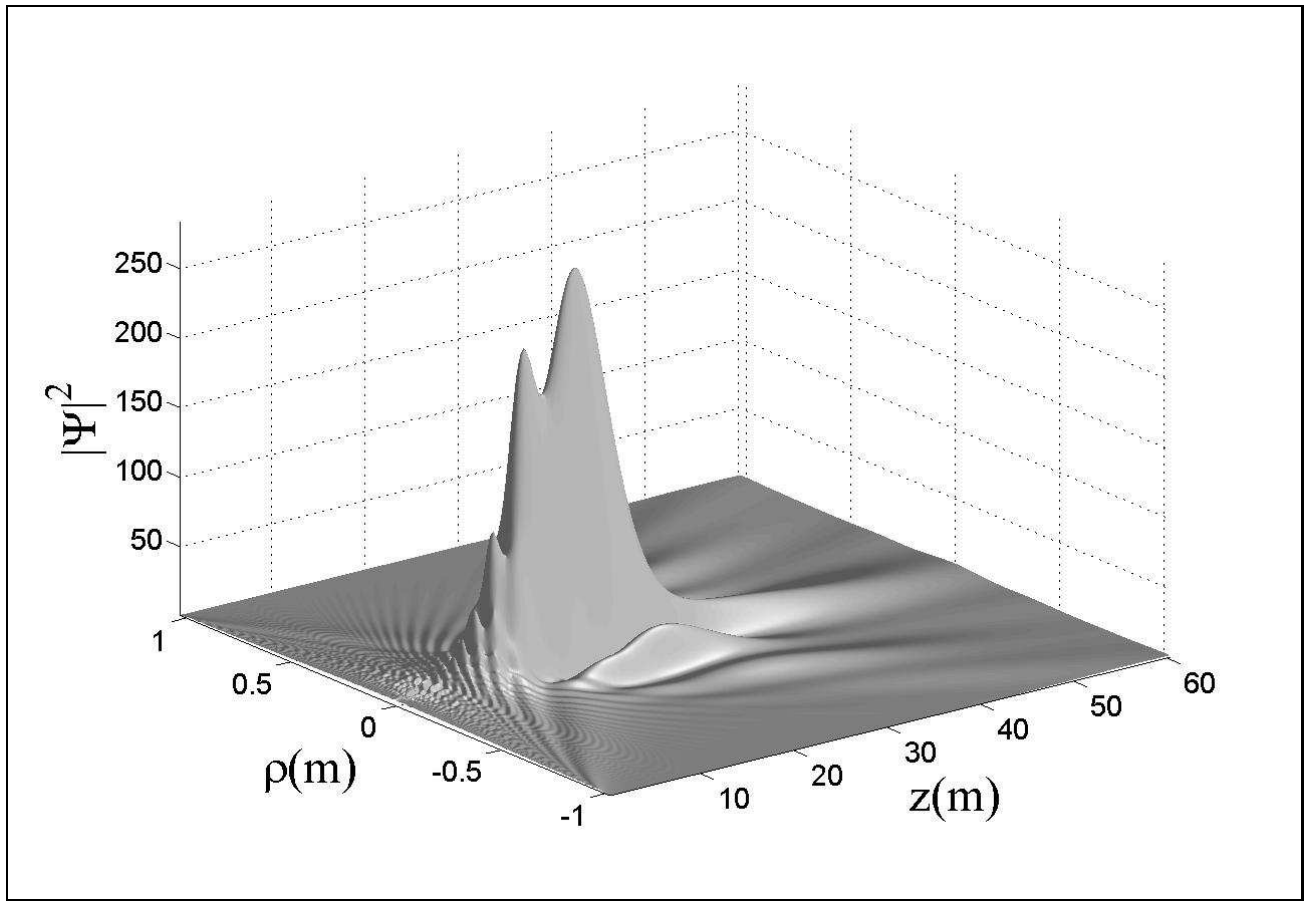}}
\caption{(a) Evolution of the beam intensity resulting from the
    paraboloid setup when $\lambda = 1$cm, $z_f = 0.5$m and $z_p = 0.525$m.}
\label{fig8}
\end{figure}

\begin{figure}[htbp]
\centering \fbox{\includegraphics[width=\linewidth]{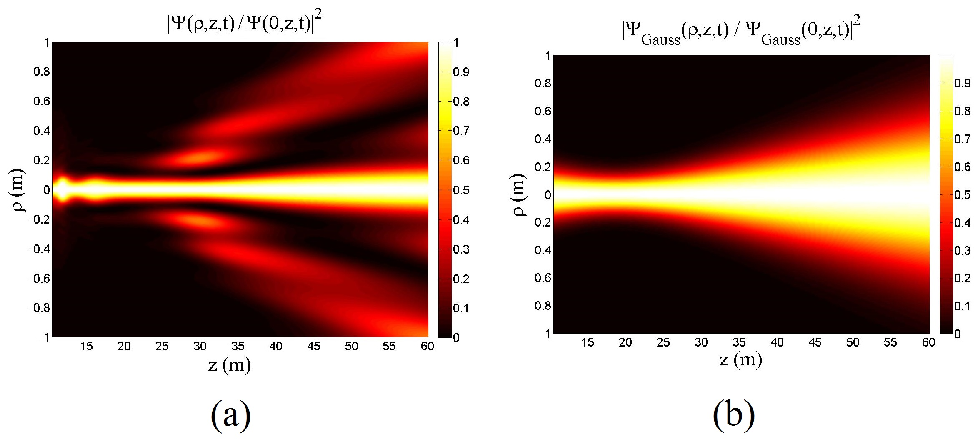}}
\caption{(a) Intensity ratio $|\Psi(\rho,z,t)/\Psi(\rho=0,z,t)|^2$
for the resulting beam starting from $z=D_{min}=10.5$m. The beam
demonstrates resistance to the diffraction effects for distances
even greater than $Z \approx 42.5$m; (b) The same kind of
intensity ratio for a Gaussian beam with the same waist at the
same position. After a distance of $17$m this beam doubles its
intensity spot size} \label{fig9}
\end{figure}

\h Figure 8 shows the beam intensity and we can see that the
extended focus occurs as predicted.

\h As before, to get more information about the transverse beam
behavior, Fig.9a shows the intensity ratio
$|\Psi(\rho,z,t)/\Psi(\rho=0,z,t)|^2$ for the resulting beam
starting from $z=D_{min}=10.5\text{m}$. Except for a small initial
oscillation of the spot radius, the evolution of the spot size
within the extended focus range is satisfactorily described by
Eq.(\ref{spot}) with eqs.(\ref{eq8},\ref{eq9}), being the size and
position of the beam waist in agreement with the predicted values.
The beam demonstrates resistance to the diffraction effects until
the distance $Z \approx 42.5$m. Actually its central spot suffers
just small variations for distances greater than the predicted
field depth $Z$.

\h Figure 9b shows the same kind of intensity ratio for a Gaussian
beam with the same waist at the same position. After a distance of
$17$m this beam doubles its intensity spot size.

\h The above examples in the optical and microwave frequency
ranges confirm our predictions about the possibilities of
generating diffraction resistant beams through using a parabolic
reflector and a source of spherical waves slightly displaced from
its focus.

\section{Conclusion}

\h This paper presents a simple way to generate diffraction
resistant scalar beams by using as hypothetical experimental setup
a parabolic reflector and a source of spherical waves located at a
position slightly displaced away from the focus of the reflector.

\h We have obtained a set of approximative equations describing
some important characteristics of the new beams, like the extended
focus range, the evolution of the beam spot size, the diffraction
resistance length, etc.. We also have obtained the respective
Fresnel diffraction integral which describes the resulting beams
and two examples have been provided, in optical and microwave
frequencies, confirming our predictions.

\h Due to its simplicity and low-cost characteristic, the method
here presented can be a very interesting alternative for the
generation of long range diffraction resistant beams in optics and
microwave applications.

\h A more complete and rigorous analysis of the setup proposed
here, taking into account the vectorial nature of the fields, will
be addressed in a future paper.

\h This work was supported by FAPESP (under grants 2013/26437-6
and 2014/04867-1); CNPq (under grant 312376/2013-8) and CAPES.

\h The authors thank Erasmo Recami, Ioannis M. Besieris and
Massimo Balma for valuable discussions and kind collaboration.


\begin{thebibliography}{11}

\bibitem{ref1}
H.~E.~Hern\'andez-Figueroa, M.~Zamboni-Rached and E.~Recami,
\emph{Localized Waves}, 1st~ed.\hskip 1em plus
  0.5em minus 0.4em\relax Hoboken, USA: John Wiley \& Sons, 2008.

\bibitem{ref2}
J.~Durnin, J.~J.~Miceli and J.~H.~Eberly, \emph{Diffraction-free
beams}, Phy.Rev.Lett.\hskip 1em plus
  0.5em minus 0.4em\relax \textbf{58},~1499-1501, 1987.

\bibitem{ref3}
I.~S.~Gradshteyn and I.~M.~Ryzhik, \emph{Table of Integrals,
Series and Products}, 5th~ed.\hskip 1em plus
  0.5em minus 0.4em\relax USA: Academic Press, 1996.

\bibitem{ref4}
J.~Goodman, \emph{Introduction to Fourier Optics}, 2nd~ed.\hskip
1em plus
  0.5em minus 0.4em\relax USA: McGraw-Hill, 1996.

\bibitem{ref5}
M.~C.~Assis, \emph{Parabolic reflectors used in generation of
non-diffracting beams}, M.Sc.~thesis.\hskip 1em plus
  0.5em minus 0.4em\relax Campinas State University, 2013.

\bibitem{ref6}
J.~W.~M.~Baars, \emph{The Paraboloidal Reflector Antenna in Radio
Astronomy and Communication}, 1st~ed.\hskip 1em plus
  0.5em minus 0.4em\relax USA: Springer, 2007.

\bibitem{ref7}
Michel Zamboni-Rached and Erasmo Recami ``Parabolic antennas, and
circular slot arrays, for the generation of Non-Diffracting Beams
of Microwaves,'' arXiv:physics/1408.5635 vl, Aug. 24, 2014.


\bibitem{ref8}
L.~A.~Ambr\'osio, \emph{Localized beams in optical tweezers with
conventional and metamaterial particles}, Ph.D~thesis.\hskip 1em
plus
  0.5em minus 0.4em\relax Campinas State University, 2010.

\bibitem{ref9}
L.~Stanislav, \emph{On the generation and application of localized
waves}, M.Sc.~thesis.\hskip 1em plus
  0.5em minus 0.4em\relax Virginia Polytechnic Institute, 2001.

\bibitem{ref10}
M.~Zamboni-Rached, \emph{Ondas localizadas aplicadas aos meios
difrativos/dispersivos}, Ph.D~thesis.\hskip 1em plus
  0.5em minus 0.4em\relax Campinas State University, 2004

\bibitem{ref11}
F.Gori, G. Guattari and C.Padovani, \emph{Bessel-Gauss Beams},
Optics Communications.\hskip 1em plus
  0.5em minus 0.4em\relax \textbf{64},~ 491-495, 1987.

\bibitem{ref12}
  Michel Zamboni-Rached, Leonardo A. Ambr\'osio, and Hugo E. Hern\'andez-Figueroa,
  ``Diffraction-attenuation resistant beams: their higher order versions and finite-aperture
  generations,'' Appl. Opt. 49, 5861-5869 (2010).

\bibitem{ref13}
  G. A. Siviloglou and D. N. Christodoulides, ``Accelerating finite energy Airy beams,''
  Opt. Lett. 32, 979-981 (2007)

\bibitem{ref14}
  Michel Zamboni-Rached, K. Z. N\'obrega, and C. A. Dartora, ``Analytic description of Airy-type
  beams when truncated by finite apertures,'' Opt. Express 20, 19972-19977
  (2012).

 \bibitem{ref15}
 Ioannis M. Besieris and Amr M. Shaarawi, ``Accelerating Airy beams with non-parabolic
  trajectories,'' Optics Communications, Vol. 331, 235-238 (2014).

\bibitem{ref16}
I. M. Besieris and A. M. Shaarawi, ``Localized monochromatic and
pulsed waves in hyperbolic metamaterials (invited paper),''
Progress In Electromagnetics Research, Vol. 143, 761-771, 2013.

\end{thebibliography}
\end{document}